# Aging effect in Magnetotransport Property of Oxygen adsorbed BaFe$_2$As$_2$

Nilotpal Ghosh, Santhosh Raj

*School of Advanced Sciences, VIT University, Chennai Campus-600127, Tamilnadu, India*
*\*Email: ghosh.nilotpal@gmail.com, nilotpal@vit.ac.in*

**Abstract.** Presence of Oxygen ($O_2$) has been found by Energy Dispersive X-ray Analysis (*EDAX*) on the surfaces of flux grown BaFe$_2$As$_2$ single crystals which were kept in air ambience for several months. Transport studies show that the $O_2$ adsorbed crystals are more resistive and do not display any sharp slope change near 140 K which is the well known Spin Density Wave (*SDW*) transition temperature ($T_{SDW}$) accompanying structural transition for as grown BaFe$_2$As$_2$. An anomalous slope change in resistivity is observed around 18 K at 0 and 5T. Magnetoresistance (*MR*) is noticed to increase as a function of applied field (*H*) quite differently than that for as grown crystals below $T_{SDW}$ which may be attributed to aging effect.

## INTRODUCTION

BaFe$_2$As$_2$ is the well studied parent compound in the 122 family of unconventional FeAs based superconductors which undergoes Spin Density Wave (*SDW*) transition around 140 K ($T_{SDW}$) accompanied by tetragonal (paramagnetic) to orthorohmbic (antiferromagnetic) structural transition [1]. Strong influence of adsorbed oxygen (air) in polycrystalline BaFe$_2$As$_2$ to mask the spin state of Fe is reported [2]. Linear magnetoresistance (*MR*) is observed in BaFe$_2$As$_2$ and attributed to Dirac fermions [3]. However, it is reported that non linear *MR* is also observed for BaFe$_2$As$_2$ single crystals which are annealed with BaAs powder [4]. Recently large low temperature *MR* is noticed for SrFe$_2$As$_2$ single crystals which is attributed to sample aging effect [5]. So far, not much work has been done in this area. Here, we report the results of our magnetotransport measurements on oxygen ($O_2$) adsorbed and aged BaFe$_2$As$_2$ single crystals i.e BaFe$_2$As$_2$:$O_2$ with as grown BaFe$_2$As$_2$.

## EXPERIMENTS

Single crystals of BaFe$_2$As$_2$ are grown by the self flux, where excess FeAs is used as a flux [6]. The crystals were kept in air ambience for several months to study the effect of oxygen adsorption and aging. Composition of the crystals has been determined by Energy Dispersive X-ray Analysis *(*EDAX*)*. Temperature dependent *MR* has been measured in the 4-probe geometry in commercial PPMS system by Cryogenic Limited at 0 and 5T magnetic fields. In order to investigate the field dependence of *MR*, measurements have been carried out at various constant temperatures below $T_{SDW}$ by ramping the magnetic field up to 10T.

## RESULTS AND DISCUSSIONS

Figure1 describes the results of EDAX on BaFe$_2$As$_2$:$O_2$. The intensity peak for oxygen is clearly observed. The oxygen is adsorbed on the surface of as grown BaFe$_2$As$_2$ crystals due to exposure in air for several months. Adsorbed $O_2$ is known to be an effective electron acceptor forming $O_{2-}$ which can permeate the crystal and form spin clusters around Fe. As a result, Fe spin state gets effected [2]. Resistivity as a function of temperature is measured for as grown BaFe$_2$As$_2$ in 4 probe geometry which shows clear magnetostructural transition at 140K ($T_{sdw}$) as reproduced in Figure 2. After the sample was left in air ambience for several months the resistivity is increased and there is no sharp slope

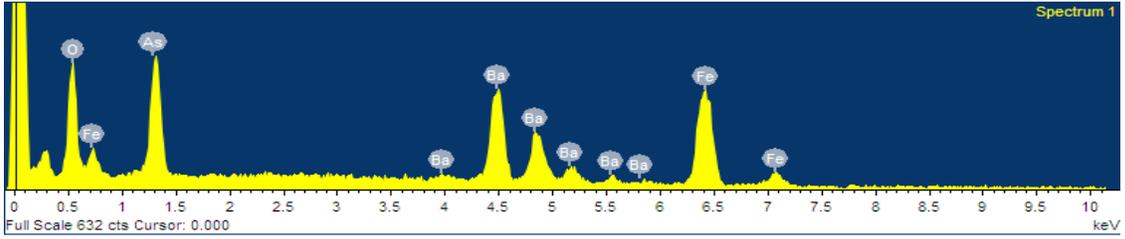

**FIGURE1.** The results of EDAX measurements of $BaFe_2As_2$ single crystals which were kept in air ambience for several months.

change around $T_{sdw}$. However, an anomalous slope change occurs at 18 K in resistivity following a down turn.

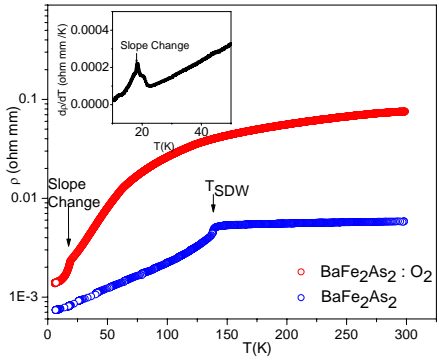

**FIGURE 2.** The results of resistivity vs temperature measurement for $BaFe_2As_2$:$O_2$ and $BaFe_2As_2$. Inset shows $d\rho/dT$ vs $T$ for $BaFe_2As_2$:$O_2$.

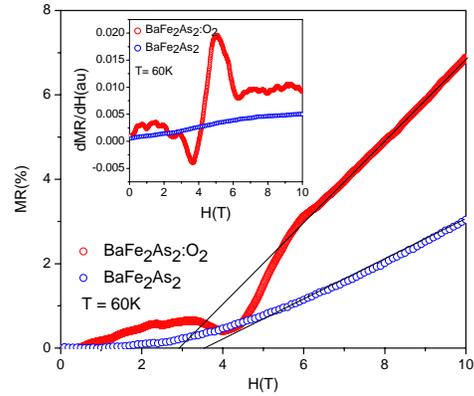

**FIGURE 4.** $MR$ vs $H$ for $BaFe_2As_2$:$O_2$ and $BaFe_2As_2$, shown in positive direction, displayed at a representative temperature 60K. The straight line guides the eye for the linear region. The inset shows $dMR/dH$ vs $H$ for both.

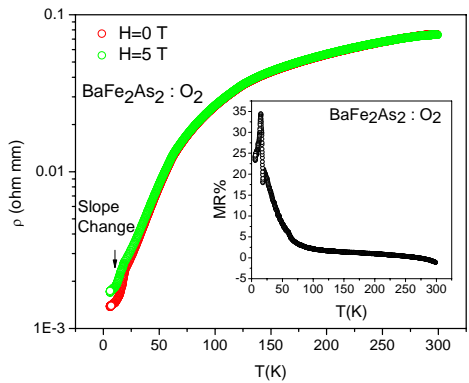

**FIGURE 3.** The results of resistivity vs temperature measurement for $BaFe_2As_2$:$O_2$ at $H = 0$ and 5 T. The inset shows MR as a function of temperature.

The resistivity when measured under magnetic field $H = 5T$, the anomalous slope change still remains (see Figure 3). The down turn in resistivity indicates the possible presence of an inhomogeneous electronic state where superconducting and non-superconducting regions do exist simultaneously but superconducting fraction may be below percolation threshold [5]. $MR$ [$=\rho(H)-\rho(0)/\rho(0)$] is found to be maximum as 35% around 15 K and negligible above $T_{sdw}$ ( inset of Figure 3). $MR$ measured as a function of $H$ up to $H = 10T$ for $BaFe_2As_2{:}O_2$ at various constant temperatures below $T_{SDW}$ and results are reproduced for a representative temperature 60 K in Figure 4. The $MR$ appears to be linear above $H = 6T$, but below it has a nonlinear $H$ dependence. In contrast, $MR$ for as grown $BaFe_2As_2$ has linear and parabolic field dependence at high and low $H$ at 60 K due to quantum and classical transport respectively [3]. Inset of Figure 4 shows $dMR/dH$ vs $H$ for $BaFe_2As_2{:}O_2$ and. $BaFe_2As_2$ single crystals where the difference is clearly realized. It is to be considered that $BaFe_2As_2{:}O_2$ crystals have strains and defects induced by aging effect. In addition, there may be spin clusters present due to $O_2$ adsorption. Hence, the situation is more complicated in case of $BaFe_2As_2{:}O_2$ crystals which can be attributed to its nonlinear and peculiar $MR$ response.

In conclusion, we have demonstrated the effect of aging and $O_2$ adsorption on the magnetotransport property of $BaFe_2As_2$ single crystals which is quite interesting and worth exploring further.

## ACKNOWLEDGMENTS

NG and SR want to thank DST-SERB, Government of India under project SR/S2/CMP-0123/2012 for financial support. NG wants to thank UGC-DAE Consortium for Scientific Research, Kolkata for providing facility for magnetotransport measurement.